\documentclass[conference]{IEEEtran}
\usepackage{cite}
\usepackage{graphicx}
\graphicspath{ {images/} }
\usepackage{graphicx}
\usepackage{multirow}
\usepackage{booktabs}
\usepackage{verbatim}
\usepackage{lscape}

\begin{document}
\bstctlcite{IEEEexample:BSTcontrol}


\title{Speech Signal Analysis for the Estimation of Heart Rates Under Different Emotional States}


\author
{\IEEEauthorblockN{Aibek Ryskaliyev, Sanzhar Askaruly}
\IEEEauthorblockA{School of Engineering\\
Nazarbayev University, Astana\\
Email: aryskaliyev@nu.edu.kz, saskaruly@nu.edu.kz }
\and
\IEEEauthorblockN{Alex Pappachen James, \it{IEEE Senior Member}}
\IEEEauthorblockA{School of Engineering, Nazarbayev University\\
www.biomicrosystems.info\\
Email: apj@ieee.org }
}
\maketitle


\begin{abstract}
A non-invasive method for the monitoring of heart activity can help to reduce the deaths caused by heart disorders such as stroke, arrhythmia and heart attack. Human voice can be considered as
a biometric data that can be used for estimation of heart rate. In this paper, we propose a method for estimating the heart rate from human speech dynamically using voice signal analysis and by development of a empirical linear predictor model. The correlation between voice signal and heart rate are established by classifiers and prediction of the heart rates with or without emotions are done using linear models. The prediction accuracy was tested using the data collected from 15 subjects, it is about 4050 samples
of speech signals and corresponding electrocardiogram samples. The proposed approach can use for early non-invasive detection of heart rate changes that can be correlated to emotional state of the individual and also can be used as tool for diagnosis of heart conditions in real-time situations.

\end{abstract}

\begin {IEEEkeywords}
Speech, Heart rates, Emotions, Telemedicine
\end{IEEEkeywords}

\section{Introduction}
{H}{eart} diseases that include cerebral stroke and other cardiovascular disorders are one of the major mortality causes in the world \cite{nelson2007physical, kotsiantis2007supervised, shahzad2013feature}. The investments in preventive healthcare such as using modern monitoring tools and devices, can help with reducing the costs of treatment and progression of the poor health conditions. This paper analyzes the existing methodologies for contact less heart rate measurement methods, and proposes more robust technique for the estimation of heart rate from the 
voice recordings using classification and prediction models \cite{fraley1998many, meraoumia2012multimodal, staelin2003parameter}.

In this work, voice signal is considered to contain biometric data reflective of the physical condition of the heart  e.g. heart rate index. It is shown in \cite{schuller2013automatic,kaur2014extraction,james2015heart, sakai2015modeling, mesleh2012heart, ruiz2013investigating} that there is an implicit relationship between heart rates and human voice. Further, there are studies that indicates a direct relation of  psychological state, say emotion to that of  heart rate and voice \cite{sakai2015feasibility}.  This is due to the fact that there are many blood vessels and capillaries in the vocal tract, which affect the voice spectrum \cite{chauhan2014mel}.

This work will give technical and algorithmic insights, including statistical approach, which could be a platform for the further advance of this research. The design of the general model, which includes
the classification and prediction model, is the primary focus of this work. Finally, the results and discussion part of the work is provided. The analysis and comparison of different approaches of the 
prediction models are given.

\subsection{Significance}

Consistent monitoring of heart rate and blood pressure is essential for people who have cardiac disorders.  In order to reduce the risk of the severity of the consequences resulting from the disease, the process of heart's physiological monitoring has to be made more convenient and affordable. Contact-less monitoring is proposed as an alternative solution to monitoring health disorders as part of a standard telehealth system.


\section{Literature Review}

There are many studies conducted on the presentation of implicit or any other means of 
relationship between heart rate and human voice. This section will briefly summarize what has been accomplished in the field by providing constructive critique on what could have been addressed from a different 
perspective.

In \cite{hermansky1990perceptual}, the author proposes a method to estimate blood pressure (BP) by applying voice-spectrum analysis. The author claims that if the results of the experiment are accurate 
enough, it could be feasible to estimate blood pressure from the recorder of the mobile device. The turning point in his work is the calculation of correlation coefficient between voice-spectrum and blood 
pressure. However, there are aspects of the work that could be addressed more accurately \cite{hermansky1991compensation}. For instance, the data collection process could have been approached more 
effectively if the recording of the voice and blood pressure measurement were performed simultaneously; but not like it was described in the original work; since the recording process of the voice and 
measuring of blood pressure separately might lead to inconclusive results. Additionally, the number of subjects seems to be very few, only two people. It barely can serve as the support to the author's main 
claims, \cite{singhal1997dynamic}, \cite{zhang2004optimality}, \cite{kim2006some}, and \cite{kibriya2004multinomial}.

In \cite{hermansky1995auditory}, the authors extract heart rate parameters, performing analysis on speech signals, in order to establish relationship between them. There is an assumption from the author's 
perspective that the heart rate can be represented as a function of several statistical parameters such as entropy, mean, and energy \cite{hermansky1992rasta}. Although, there is a correlation it is still
unclear why these parameters were chosen. The researcher did not qualify what was the basis for completing that particular statistical analysis and which parameters constituted the changes in the heart rate \cite{meseguer2009speech}.

In \cite{chauhan2014mel, ayu2012comparison}, researchers showed that there is indeed a correlation between heart rate and voice signal. They found a group of classes which could serve as an index of 
relationship between physiological parameters such as heart rate, skin conductance and voice signal. There had been different machine learning algorithms, namely, Support Vector Regression (SVR), Support 
Vector Machines (SVM), Sequential Minimal Optimization (SVR), and binary classification models, implemented. Additionally, there are many Low Level Descriptors (LLD) for voice signal feature extraction, and 
statistical functions, which are available in openSMILE software \cite{hoffman2002novel,breiman1996bagging}. However, they do not explicitly show, which feature extraction methods and 
algorithms are constituting to the estimation of factors between HR and human speech, \cite{stern2001psychophysiological}.

\section{Methodology}

\subsection{Data Collection}
  Heart rate via ECG device along with the speech signal was measured and 
recorded. Totally, 15 subjects participated in the data collection process. They were instructed to pretend as they were experiencing three types of emotions, namely, joy, neutral and anger, in different 
situations respectively. Overall 90 samples of heart rate and speech signal for each emotion for each subject were collected. It is about 4050 samples in total. All the analysis performed in this paper assume and take into account the dynamic nature of the heart rates so as to reflect a real-time application.
\subsection{Feature Extraction}
The next step is to extract the heart rate from the ECG samples, and calculate the feature distance estimates from the speech signals. Heart rates were estimated by 1500 rule and feature distances were extracted by applying Fourier Transform and Mel-frequency cepstral coefficients (MFCC) coefficients. 
\subsection{Data Filtering}
The aforementioned procedure was followed by the revision of the data for the presence of any errors or mistakes associated with the human factor, i.e. mistyping. The irrelevant or impossible values of 
heart rate and feature distances were immediately discarded from the dataset.

The next step was to develop the classification and prediction model. In order to find out whether the classification is necessary, the prediction model for the combined emotions data and separate emotions data was 
done. Eventually, the classification of emotions was the essential part for the estimation of correlation between heart rate and feature distance. This was due to the fact that the prediction of separate 
emotions showed higher accuracy than the prediction for the combined emotions.

\subsection{Classification of Emotions}

Several classifications are provided. The most accurate is the algorithm named Classification via Regression. This algorithm applies a method of building a classification tree for the actual 
classification procedure. The output of the algorithm is assigned a new value, in our case is the emotion, joy, neutral or anger, for corresponding heart rates for each of the subjects. By the set of the 
feature distances provided, each of the observations will get to one of the terminal nodes of the tree. New observation is assigned a class of emotion of the terminal node, where the output belongs to. In the 
Table 4.4 of the next section, different classifications algorithms and the corresponding accuracy rates are shown. It can be seen that the 
Classification via Regression is the most accurate and this algorithm is chosen for the construction of the general model.

\subsection{Prediction Model}

The prediction of heart rate values in beats per minute from the given feature distance values is necessary. There is indeed a correlation between these two physiological factors; however, it is 
non-deterministic in its nature. In other words, there is no explicit relationship but rather an implicit one. In order to establish the relation between heart rate and feature distance the empirical model 
must be built. The regression analysis is applied for that purpose. The heart rate and feature distance values indicated an approximate linear relationship, and supported towards the use of linear regression model.

\subsection{Theory: Simple Linear Regression}

The idea behind the regression analysis is simple. The regression line is fitted in the dataset of $n$ points. The function that describes the relationship of $x$ and $y$ is given in Eqn. \ref{eq:regression}, 
where $x$ is an independent or predictor variable, and $y$ is a dependent or response variable. In our case, the feature distance (FD) is a predictor variable and heart rate (HR) is a response variable, 
$\left\{(x_i,~y_i),~i = 1,~2,~...~,~n\right\}$.

\begin{equation} \label{eq:regression}
    y_i = \beta_0 + \beta_1 x_i + \epsilon_i
\end{equation}

In Eqn. \ref{eq:regression}, $\beta_0$ is the $y-intercept$ term of the true line, $\beta_1$ is the slope of the true line, and $\epsilon_i$ is the random error component of the true line. It is important to 
mention that a true line is the actual line that fits all of the data points. However, it is impossible to estimate in practice. Thus, there is another concept of the line is being introduced. This is the 
regression line. The regression line is given by the Eqn. \ref{eq:reg}.

\begin{equation} \label{eq:reg}
    \hat{y_i} = \hat{\beta}_0 + \hat{\beta}_1x_i
\end{equation}

In the Eqn. \ref{eq:reg}, $\hat{y}_i$ is the predicted value of the $i$th index, $\beta_0$ and $\beta_1$ are the estimators of the parameters of $\beta_0$ and $\beta_1$ as in Eqn. \ref{eq:regression}. Finally, the Eqns. \ref{eq:beta1} and \ref{eq:beta0}, show the least square estimators for the simple linear regression model.

\begin{equation} \label{eq:beta1}
    \hat{\beta}_1 = \frac{S_{xy}}{S_{xx}}
\end{equation}

\begin{equation} \label{eq:beta0}
    \beta_0 = \bar{y} - \hat{\beta}_1\bar{x}
\end{equation}

In the Eqn. \ref{eq:beta1}, $S_{xy} = \sum_{i=1}^{n}(x_i-\bar{x})(y_i-\bar{y})$ and $S_{xx} = \sum_{i=1}^{n}(x_i-\bar{x})^2$.

\subsection{Theory: Other Mathematical Formulas used in this work}

In the Eqn. \ref{eq:mean}, $\bar{x}$ is the sample mean. The corresponding population mean is denoted by $\nu$.

\begin{equation} \label{eq:mean}
    \bar{x} = \frac{1}{n}\sum_{i=1}^n x_i
\end{equation}

\subsubsection{Sample Mean and Population Mean}

The $variance$ of a sample of measurements $x_1,~x_2,~...~,~x_n$ is the sum of the square of the differences between each measurement value and their mean, divided by $n-1$. The equation is given in Eqn. 
\ref{eq:var}.

\begin{equation} \label{eq:var}
    s^2 = \frac{1}{n-1}\sum_{i=1}^{n}(x_i-\bar{x})^2
\end{equation}

\subsubsection{Variance and Standard Deviation}

The $standard~deviation$ of a sample of measurements is the positive square root of the variance, as shown in Eqn. \ref{eq:dev}. The corresponding $population~standard~deviation$ is provided in Eqn. 
\ref{eq:pop}.

\begin{equation} \label{eq:dev}
    s = \sqrt{s^2}
\end{equation}

\begin{equation} \label{eq:pop}
    \sigma = \sqrt{\sigma^2}
\end{equation}

\subsubsection{Relative Error Estimation}

In the Eqn. \ref{eq:error}, the relative error estimation formula is given. $HR_{estimated}$ is the heart rate estimated from the prediction and $HR_{measured}$ is the subject's actual heart rate that was 
extracted from electrocardiogram sample in beats per minute.

\begin{equation} \label{eq:error}
    Relative~Error = \frac{|HR_{estimated} - HR_{measured}|}{HR_{measured}}\times100
\end{equation}

\subsubsection{The Normal Distribution, Gaussian Curve}

\begin{figure}[ht!]
    \centering
    \includegraphics[scale=.45]{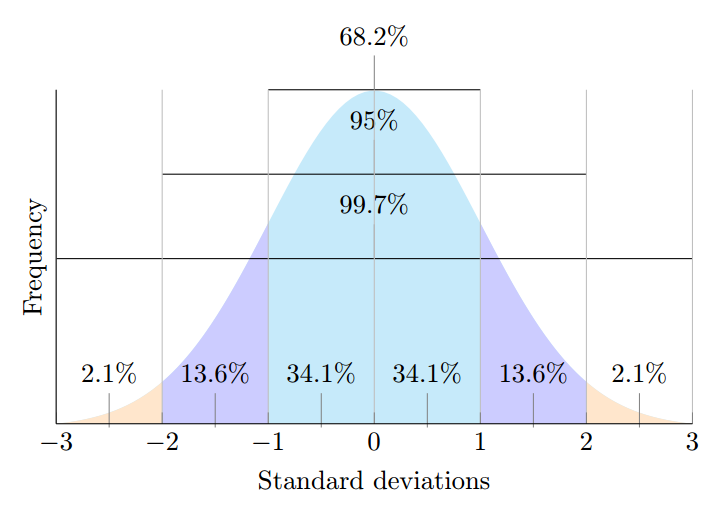}
    \caption{Normal Curve.}
    \label{fig:bell}
\end{figure}

For a distribution of measurements that is approximately normal, i.e. bell shaped, it follows that the interval with the end points, the illustration is given in Fig. \ref{fig:bell}.
	$\mu\pm\sigma$ contains approximately 68\% of the measurements, 
	$\mu\pm2\sigma$ contains approximately 95\% of the measurements,
	$\mu\pm3\sigma$ contains almost all of the measurements, \cite{hou2013efficient}.

\section{Results and Discussion}

There were total of 15 subjects participating in the data collection process. As was described earlier, in section III, the prediction model was done for two types of scenarios. Firstly, the regression analysis was done when the dataset was separated based on the emotional state of the subject. The maximum prediction accuracy among 15 subjects was estimated as 97.36\% for the joy, 97.40\% for the neutral state, and 97.76\% for the anger. The average prediction accuracy was above 90\% for all of the subjects.

Second experiment, was done almost in the same manner, except the data for each emotion for the given subject was combined. The average accuracy rate in this case was also above 90\%. However, further analysis and comparison of relative error and accuracy calculations showed that the accuracy for separate emotions was a bit higher than the combined ones, for more detailed description refer to the Comparison and Analysis of this section.

The next section gives the insights on the classification procedure. The WEKA software tool was used for this purpose. Several classification algorithms were applied including Classification via Regression, Naive Bayes Classification. Finally, the general model of the classification and prediction was done. The accuracy of the estimation was reported and it was around 60\% at most. The Classification via Regression proved to be the most robust and accurate algorithm.

\subsection{Experiment 1 -- Separate Emotions}

This experiment was primarily conducted for the estimation of accuracy rates for the prediction of heart rates, from given feature distance values, for separated dataset based on different emotions.  The Table \ref{table:separ} shows the results obtained from the accuracy rate estimation for each subject for each type of emotion. The error rates are quite low. The maximum relative prediction error is 9.20\% which indicates good regression analysis.

\begin{figure}[ht!]
    \centering
    \includegraphics[scale=.5]{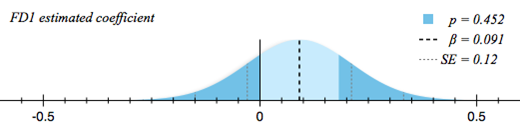}
    \caption{Estimated coefficient of feature distance (FD) for subject 1, for emotion of Joy.}
    \label{fig:joy1}
\end{figure}

\begin{figure}[ht!]
    \centering
    \includegraphics[scale=.5]{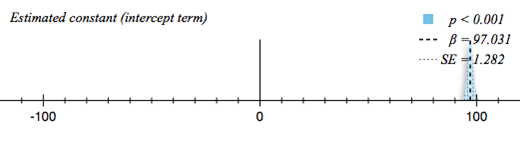}
    \caption{Estimated constant, the intercept term of feature distance (FD) for subject 1, for emotion of Joy.}
    \label{fig:joy2}
\end{figure}

 The Fig. \ref{fig:joy1} shows the example of the prediction model and its plot for the emotion of Joy, for the subject 1. This figure gives the values for the estimator of the slope, estimated coefficient, which is equal to 0.091 and the estimator for then intercept term, i.e. estimated constant, is shown in Fig. \ref{fig:joy2}, which is equal to 97.031. From the Eqn. \ref{eq:reg}, in section III, the predicted heart rate can be calculated. This was the case for each subject and their corresponding emotion dataset.

The Table \ref{table:separ}, gives the accuracy and error rates for each of the emotion of each subject that was calculated by Eqn. \ref{eq:reg}.

\begin{table}[ht!]
\centering
\caption{Relative Error and Accuracy in (\%) estimations for each subject for different emotional state.}
\label{table:separ}
\begin{tabular}{c c c c c c c}
\toprule
\multirow{2}{*}{Subject No.} & \multicolumn{3}{c}{Error} & \multicolumn{3}{c}{Accuracy} \\ 

                            & Joy    & Neutral  & Anger  & Joy     & Neutral   & Anger   \\ 
                            \midrule
1                           & 4.42   & 3.16     & 3.96   & 95.58   & 96.84     & 96.04   \\ 
2                           & 6.20   & 8.83     & 7.52   & 93.80   & 91.17     & 92.48   \\ 
3                           & 6.69   & 6.28     & 6.04   & 93.31   & 93.72     & 93.96   \\ 
4                           & 3.23   & 4.19     & 3.56   & 96.77   & 95.81     & 96.44   \\ 
5                           & 2.77   & 3.43     & 3.70   & 97.23   & 96.57     & 96.30   \\ 
6                           & 2.64   & 5.27     & 2.68   & 97.36   & 94.73     & 97.32   \\ 
7                           & 7.09   & 6.73     & 5.08   & 92.91   & 93.27     & 94.92   \\ 
8                           & 6.41   & 8.01     & 9.20   & 93.59   & 91.99     & 90.80   \\ 
9                           & 5.93   & 3.18     & 3.55   & 94.07   & 96.82     & 96.45   \\ 
10                          & 3.05   & 2.60     & 2.24   & 96.95   & 97.40     & 97.76   \\ 
11                          & 2.77   & 2.83     & 2.54   & 97.23   & 97.17     & 97.46   \\ 
12                          & 6.94   & 3.49     & 6.02   & 93.06   & 96.51     & 93.98   \\ 
13                          & 6.06   & 3.41     & 3.92   & 93.94   & 96.59     & 96.08   \\ 
14                          & 6.43   & 6.76     & 3.69   & 93.57   & 93.24     & 96.31   \\ 
15                          & 6.67   & 4.24     & 5.13   & 93.33   & 95.76     & 94.87   \\ 
\bottomrule
\end{tabular}
\end{table}

\subsection{Experiment 2 -- Combined Emotions}

The aim of the Experiment 2 was to observe whether the results will be higher if the data samples for each subject were not differentiated based on their emotional states. The Table \ref{table: gen} represents the results obtained. As it can be seen, the accuracy rates for the prediction for the emotions combined is also quite good, almost as in Experiment 1. The minimum prediction accuracy is 75.88\%.

The same procedure was done as in Experiment 1, but in this case the emotions of each subject were not 
differentiated. The Table \ref{table:  gen}, illustrate the accuracy and error rates calculated by the equation given in Eqn. \ref{eq:reg}.

\subsection{Comparison and Analysis}
The comparison of the same approach for two different scenarios is discussed in this part. The idea is simple, from the Table \ref{table: gen}, it can be seen that the relative error for the separate emotions is less compared to the error rate for the combined ones. Although the difference is not very much, it serves as the basis for the conduction of classification.

\begin{table}[ht!]
\centering
\caption{The Relative Error rates in (\%) of the prediction, including combined emotions, average of separate and separate emotions, for each subject respectively.}
\label{table: gen}
\begin{tabular}{c c c c c c}
\toprule
Subject No. & Combined & Average & Joy  & Neutral & Anger \\ 
\midrule
1          & 4.82              & 3.85                         & 4.42 & 3.16    & 3.96  \\ 
2          & 8.14              & 7.52                         & 6.2  & 8.83    & 7.52  \\ 
3          & 6.45              & 6.34                         & 6.69 & 6.28    & 6.04  \\ 
4          & 4.44              & 3.66                         & 3.23 & 4.19    & 3.56  \\ 
5          & 24.12             & 3.3                          & 2.77 & 3.43    & 3.7   \\ 
6          & 3.71              & 3.53                         & 2.64 & 5.27    & 2.68  \\ 
7          & 6.49              & 6.3                          & 7.09 & 6.73    & 5.08  \\ 
8          & 8.93              & 7.88                         & 6.41 & 8.01    & 9.2   \\ 
9          & 6.27              & 4.22                         & 5.93 & 3.18    & 3.55  \\ 
10         & 3.31              & 2.63                         & 3.05 & 2.6     & 2.24  \\ 
11         & 4.39              & 2.71                         & 2.77 & 2.83    & 2.54  \\ 
12         & 5.87              & 5.48                         & 6.94 & 3.49    & 6.02  \\ 
13         & 7.57              & 4.46                         & 6.06 & 3.41    & 3.92  \\ 
14         & 6.13              & 5.63                         & 6.43 & 6.76    & 3.69  \\ 
15         & 5.65              & 5.34                         & 6.67 & 4.24    & 5.13  \\ 
\bottomrule
\end{tabular}
\end{table}

\subsection{Classification of Emotions}

 In Tables \ref{table: subj7}  and \ref{table: aver} few algorithms and their classification accuracy for each subject are given.



\begin{table*}[ht!]
\centering
\caption{Classification accuracy rates for different classification techniques for each subject, for subjects 1 - 15, where 66\% is taken as a training, and the remaining is taken as a testing.}
\label{table: subj7}
\begin{tiny}

\begin{tabular}{c c c c c c c c c c c c c c c c}
\toprule
\multirow{2}{*}{Classifier type} & \multicolumn{10}{c}{Classification Accuracy for Subject No.} \\ 
                                 & 1      & 2      & 3      & 4      & 5      & 6      & 7      & 8      & 9     & 10    & 11       & 12       & 13       & 14       & 15\\
\midrule
bayes.BayesNet                         & 58.89  & 51.11  & 58.24  & 68.13  & 81.32  & 51.65  & 33.71  & 50.00  & 72.53  & 57.14  & 89.01   & 45.65   & 69.57   & 57.14   & 43.96\\ 
bayes.NaiveBayes                         & 71.11  & 66.67  & 62.64  & 63.74  & 80.22  & 56.04  & 42.70  & 45.65  & 80.22  & 59.34  & 85.71   & 51.09   & 72.83   & 59.34   & 56.04\\ 
bayes.NaiveBayesMultinomial                         & 42.22  & 54.44  & 51.65  & 52.75  & 64.84  & 52.75  & 46.07  & 35.87  & 82.42  & 50.55  & 68.13   & 46.74   & 43.48   & 43.96   & 53.85\\  
functions.Logistic                         & 71.11  & 66.67  & 58.24  & 67.03  & 78.02  & 60.44  & 51.69  & 45.65  & 82.42  & 50.55  & 68.13   & 46.74   & 43.48   & 43.96   & 53.85\\ 
functions.MultilayerPerceptron                         & 63.33  & 62.22  & 51.65  & 69.23  & 74.73  & 56.04  & 44.94  & 42.39  & 80.22  & 59.34  & 87.91   & 53.26   & 75.00   & 58.24   & 45.05\\ 
functions.SimpleLogistic                         & 71.11  & 66.67  & 58.24  & 68.13  & 76.92  & 57.14  & 51.69  & 44.57  & 80.22  & 54.95  & 87.91   & 46.74   & 72.83   & 54.95   & 53.85\\ 
lazy.IBk                         & 57.78  & 56.67  & 53.85  & 61.54  & 75.82  & 51.65  & 35.96  & 45.65  & 81.32  & 54.95  & 87.91   & 53.26   & 75.00   & 56.04   & 46.15\\ 
lazy.KStar                         & 65.56  & 56.67  & 57.14  & 71.43  & 79.12  & 52.75  & 41.57  & 46.74  & 78.02  & 56.04  & 83.52   & 45.65   & 65.22   & 53.85   & 40.66\\ 
lazy.LWL                         & 56.67  & 65.56  & 50.55  & 62.64  & 72.53  & 50.55  & 51.69  & 48.91  & 67.03  & 56.04  & 91.21   & 43.48   & 72.83   & 60.44   & 47.25\\ 
meta.Bagging                        & 60.00  & 52.22  & 60.44  & 69.23  & 81.32  & 49.45  & 43.82  & 48.91  & 81.32  & 59.34  & 87.91   & 50.00   & 72.83   & 49.45   & 49.45\\ 
meta.ClassificationViaRegression                        & 64.44  & 66.67  & 59.34  & 68.13  & 78.02  & 58.24  & 56.18  & 45.65  & 82.42  & 57.14  & 89.01   & 53.26   & 73.91   & 59.34   & 48.35\\ 
meta.CVParameterSelection                        & 33.33  & 33.33  & 32.97  & 32.97  & 32.97  & 32.97  & 33.71  & 32.61  & 32.97  & 32.97  & 32.97   & 33.70   & 33.70   & 32.97   & 32.97\\ 
meta.CVParameterSelection                        & 56.67  & 51.11  & 58.24  & 68.13  & 81.32  & 51.65  & 33.71  & 50.00  & 73.63  & 57.14  & 89.01   & 45.65   & 70.65   & 57.14   & 43.96\\ 
rules.JRip                        & 60.00  & 65.56  & 59.34  & 65.93  & 80.22  & 52.75  & 42.70  & 48.91  & 76.92  & 56.04  & 52.75  & 42.70  & 48.91  & 76.92  & 56.04\\ 
trees.J48                        & 57.78  & 62.22  & 62.64  & 64.84  & 79.12  & 52.75  & 46.07  & 48.91  & 81.32  & 59.34  & 52.75  & 46.07  & 48.91  & 81.32  & 59.34\\ 
max accuracy percentage                        & 71.11  & 66.67  & 62.64  & 71.43  & 81.32  & 60.44  & 56.18  & 50.00  & 82.42  & 59.34  & 60.44  & 56.18  & 50.00  & 82.42  & 59.34\\ 
min error percentage                        & 28.89  & 33.33  & 37.36  & 28.57  & 18.68  & 39.56  & 43.82  & 50.00  & 17.58  & 40.66  & 39.56  & 43.82  & 50.00  & 17.58  & 40.66\\
\bottomrule
\end{tabular}
\end{tiny}
\end{table*}

\begin{table}[ht!]
\centering
\caption{Average Classification Accuracy in (\%) for all 15 subjects, for different classification algorithms.}
\label{table: aver}
\begin{tabular}{c c}
\toprule
Classifier type             & Average Accuracy \\
\midrule
bayes.BayesNet          & 59.20 \\ 
bayes.NaiveBayes          & 63.56 \\ 
bayes.NaiveBayesMultinomial          & 52.65 \\ 
functions.Logistic          & 63.86 \\ 
functions.MultilayerPerceptron          & 61.06 \\ 
functions.SimpleLogistic          & 63.27 \\ 
lazy.IBk          & 57.46 \\ 
lazy.KStar          & 61.72 \\ 
lazy.LWL          & 59.45 \\ 
meta.Bagging         & 61.34 \\ 
meta.ClassificationViaRegression         & 64.01 \\ 
meta.CVParameterSelection         & 33.14 \\ 
meta.CVParameterSelection         & 59.20 \\ 
rules.JRip         & 60.91 \\ 
trees.J48         & 61.50 \\ 
Max         & 64.01 \\ 
Min         & 35.99 \\ 
\bottomrule
\end{tabular}
\end{table}

\subsection{Classification and Prediction General Model}

The general model is very simple. The accuracy rate of the prediction model is multiplied by the accuracy of the classification. The Table \ref{table: aver} illustrates the average accuracy of each particular classification algorithm for each emotion multiplied by the simple linear regression prediction. 

\section{Conclusion}

In this work, we demonstrated that there is a strong relationship between voice spectrum and 
heart rate. This correlation was evidenced through  
feature extraction of speech signals followed by benchmark against measured ECG data in dynamic real-time measurement situations. For the voice 
spectrum analysis, MFCC was used to extract feature distances and compared with the ground truth 
estimation of heart rate from ECG samples using 1500 rule \cite{james2015heart, aha1991instance}.



The proposed work can be a foundation for the future development of contact less heart rate monitoring devices, methods and systems. And can help as a tool for early diagnosis and treatment of the patients having acute or chronic conditions.

In the future, along with increase in the scale of the dataset in volume and versatility, it can be further benchmark to a variety of prediction and classification algorithms to improve the heart rate estimation accuracy. The proposed system can be used to build real-time system making use of existing cloud infrastructure solutions and mobile devices.

\bibliographystyle{IEEEtran}
\bibliography{references}

\end{document}